\documentclass{WileyMSP-template}
\DeclareUnicodeCharacter{2212}{-}
\usepackage{natbib}
\usepackage{textgreek}
\usepackage{amsmath}
\usepackage{xcolor}
\usepackage{mathtools}
\usepackage{placeins}

\begin{document}

\pagestyle{fancy}
\setlength{\headheight}{24pt}

\title{Analog Weight Update Rule in Ferroelectric Hafnia, using picoJoule Programming Pulses}

\maketitle


\author{Alexandre Baigol}
\author{Nikhil Garg}
\author{Matteo Mazza}
\author{Yanming Zhang}
\author{Elisa Zaccaria}
\author{Wooseok Choi}
\author{Bert Jan Offrein}
\author{Laura Bégon-Lours*}


\dedication{}

\begin{affiliations}
A. Baigol, Dr. N. Garg, M. Mazza, Y. Zhang, Prof. L. Bégon-Lours\\
Address: Integrated Systems Laboratory, Department of Information Technology and Electrical Engineering, ETH Zürich, CH-8092 Zürich, Switzerland\\
Email Address: abaigol@ethz.ch

E. Zaccaria, Dr. W. Choi, Dr. B. J. Offrein\\
Address: IBM Research Europe - Zürich, Säumerstrasse 4, 8803 Rüschlikon, Switzerland\\

\end{affiliations}



\begin{abstract}

In an effort to compete with the brain’s efficiency at processing information, neuromorphic hardware combines artificial synapses and neurons using mixed-signal circuits and emerging memories. In ferroelectric resistive weights, the strength of the synaptic connection between two neurons is stored in the device conductance. During learning, programming pulses are applied to the synaptic weight, which reconfigures the ferroelectric domains and adjusts the conductance. One strategy to lower the energy cost during the training phase is to lower the duration of the programming pulses. However, the latter cannot be shorter than the self-loading time of the resistive weights, limited by intrinsic parasitics in the circuits. In this work, ferroelectric resistive weights are fabricated using a process compatible with CMOS Back-End-Of-Line integration, based on hafnia/zirconia nanolaminates. By laterally scaling the device area under 100 \textmu m\textsuperscript{2}, the self-loading time becomes sufficiently short to enable 20 ns programming, which corresponds to a maximum of 3 picoJoules per pulse. Further, in this work, the weight update rule with 20 ns pulses is experimentally measured not only for different amplitudes but also for different initial conductance states. We find that the final weight is determined by the pulse amplitude, independent of the initial weight value.

\end{abstract}

\section{Introduction}

\justifying
With the discovery of ferroelectricity in Hafnia, this material has become an attractive contender for the realization of neuromorphic circuits, artificial neural network (ANN) accelerators, and analogue non-volatile memory technologies. Among the latter, ferroelectric memories are comparable with other commercially available technologies in terms of density, retention, and fabrication complexity \cite{hellenbrandProgressEmergingNonvolatile2024}. However, they present two key advantages: first, they can be operated with low energies (fJ read and pJ write) \cite{begon-lours_high-conductance_2021} and at ultrafast speeds, even under 1 ns \cite{dahan_sub-nanosecond_2023}. Second, in contrast with other technologies such as phase change (PCM) or filamentary (ReRAM) memories, they don't rely on ion motion or phase transitions but on an electrostatic effect, which results in high endurance (up to 10\textsuperscript{14} cycles \cite{el-masry_toward_2024}). 

Ferroelectric materials have some intrinsic switching dynamics that link the proportion of switched domains to the amplitude of the electric field and the time during which it is applied \cite{merz_domain_1954,lee_nucleationlimited_2019,antoniadis_nucleation-limited_2022}. The ferroelectric coercive bias (\textit{V\textsubscript{c}}), which represents the voltage at which the net polarization value reverses sign, is pulse width-dependent \cite{merz_domain_1954,dawber_physics_2005}, increasing with the reduction of pulse duration. Therefore, the design of operation rules for a given device requires an observation of the interplay between programming pulse parameters.  
At the array level, programming with fast pulses is not only advantageous in terms of programming speed but also in terms of energy consumption and device endurance \cite{gargEnergyconvergenceTradeTraining2025a}. 

However, ferroelectric resistive devices also present capacitive properties. Consequently, intrinsic device and circuit parasitics introduce RC delays, which might prevent high-speed operation \cite{dahan_sub-nanosecond_2023}. If the contact resistance is small compared to the device resistance, the RC delay decreases as the device area decreases \cite{menzel_ultimate_2019}. This relation is detailed later in the manuscript. Therefore, laterally scaling memristive devices has two advantages: first, it lowers the energy cost for switching. Second, it enables high-speed operation. Such scaling is not trivial, as many challenges arise at the device and at the material levels. This is most notable in polycrystalline materials as the device size approaches the grain size. In aggressively scaled hafnia-based Ferroelectric Field-Effect Transistors (FeFETs), the three-terminal configuration enables sensing of discrete domain switching through subtle changes in channel conduction \cite{mulaosmanovic_switching_2017}. This revealed a strong dependence of the device-to-device variation and yield on the lateral device scale. Further work deepened the understanding of these effects, such as an abrupt, step-like switching \cite{zhu_sensing_2023} and stochastic switching events \cite{guoExploitingFeFETSwitching2021}, and how they might be harnessed for neuromorphic implementations \cite{casalsAvalancheCriticalityFerroelectric2021}. 

Working with micrometer-scale devices relaxes these constraints. Yet, switching dynamics and conductance changes in two-terminal devices are still difficult to measure due to the small currents. 
One strategy is to reduce the ferroelectric layer thickness, which reduces the capacitive contribution to the electrical response by increasing the device conduction through current leakage \cite{begon-lours_effect_2022}. This leakage masks switching dynamics but enables the non-destructive state readout through sub-V\textsubscript{c} sweeps. This thickness reduction also increases the processing complexity and requires the usage of more advanced annealing strategies to maintain CMOS compatibility. As already observed in \cite{begon-lours_back-end--line_2024}, by using a WO\textsubscript{x} interlayer under an HfO\textsubscript{2}/ZrO\textsubscript{2} nanolaminate, ferroelectricity is successfully stabilized in 5 nm thick films and integrated in the Back-End-Of-Line of CMOS technology. As in ferroelectric tunnel junctions with a single-layer \cite{garcia2014ferroelectric} or double-layer \cite{max2020hafnia}, the two-terminal devices exhibit a change in conductance upon polarization reversal, attributed to the ferroelectric field-effect. However, thermally activated conduction shows that direct tunneling is not the dominant mechanism in the studied devices. Compared to this prior work \cite{begon-lours_back-end--line_2024}, we deposit the same functional stack on W/SiO\textsubscript{2}/Si wafers, but we modify our lithography process to reduce device areas by a factor of 100, reaching OFF-state resistances in the GOhms range. In this work, we show that in the range 4 to 100 \textmu m\textsuperscript{2}, the self-loading time of the devices is lower than 20 ns, which allows the programming energy to remain in the picoJoule range without compromising the On/Off ratio.

In neuromorphic circuits and ANN accelerators based on memristive elements, transfer and online learning rely on the device's ``depression'' and ``potentiation'' characteristics: the gradual increase or decrease of the conductance upon the application of programming pulses. The symmetry and linearity of these characteristics affect the circuit-level performances of the neural network \cite{luo_mlpneurosimv30_2019}. Existing circuit-level implementations are designed for operation with identical programming pulses \cite{gokmen_algorithm_2020,rasch_fast_2024}, relevant to mature memristive technologies such as ReRAM \cite{falcone_all--one_2025} and PCM \cite{narayanan_fully_2021}. In these devices, resistive-state switching is current-driven. Instead, in ferroelectric devices, resistive switching is field-driven. Consequently, potentiation-and-depression data sets are usually presented with an increasing amplitude scheme \cite{kangTimedependentVariabilityRRAMbased2017,sternSubNanosecondPulsesEnable2021,liFourlayer3DVertical2016}. While revealing relevant parameters, such as the On/Off ratio, operational voltage, and steepness of the switch, such a scheme does not accurately represent the device behaviour in weight transfer or error backpropagation operations. In particular, it does not reflect how the conductance change depends on the initial conductance state. This can vary from one material to another: for example, Boyn \textit{et al.} showed that epitaxial ferroelectric tunnel junctions exhibit cumulative conductance change in write/ read experiments with identical pulses \cite{boyn_learning_2017}. Another example is the work by Siannas \textit{et al.}, demonstrating accumulative conductance change in polycrystalline hafnia synaptic weights, using a train of identical pulses separated by a 100 ns interval \cite{siannas_electronic_2024}.  

Some examples of neuromorphic chips leverage the non-volatility of memristive weights and implement the weight update rule in dedicated circuits, such as the TEXEL \cite{greatorexNeuromorphicProcessorOnchip2025} or UNICO \cite{garg2024versatile, garg2024neuromorphic} processors.

To enable the adoption of ferroelectric weights, it is crucial to define weight update rules specific to this class of materials and to account for RC effects. In this work, we verify the programming conditions with respect to RC delays. We then measure the weight update rule using 20 ns pulses for different amplitudes and initial conductance states. We find that the final weight is determined by the pulse amplitude and is independent of the initial weight value, and formalize the weight update rule.

\section{Results and Discussion}
\subsection{Conduction mechanisms}
\begin{figure}[hbt!]
  \includegraphics[width=1\linewidth]{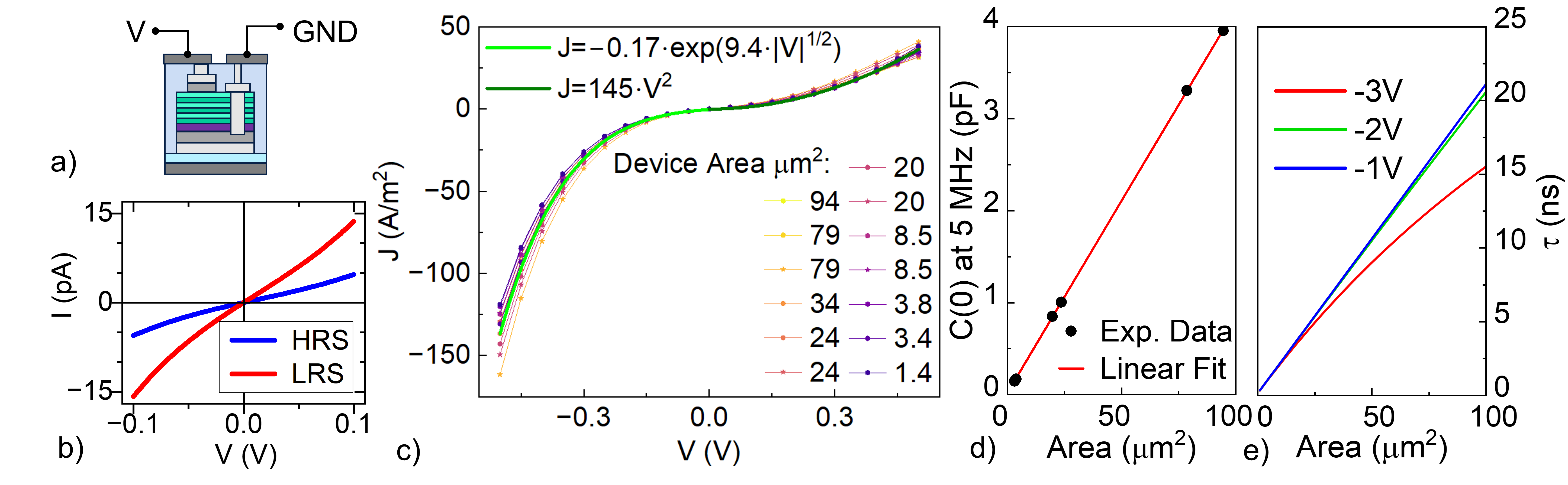}
  \caption{a) The bias is applied to the top electrode, which is used to define the device area. b) Non-destructive DC-IV in the High and Low Resistive States for a 4 \textmu m\textsuperscript{2} weight. c) During programming with negative bias, the current is electrode-limited. The DC-IV experimental data (symbols) are fitted by an exponential law (light green). During programming with positive bias, it is bulk-limited and fitted with a power law (dark green). d) Capacitance of different-sized devices measured at 5 MHz. e) The characteristic charging time (\texttau) remains below 20 ns for all the measured devices and voltage ranges.}
  \label{fig:Conduction}
\end{figure}

First, we analyze the conduction mechanisms in the fabricated devices. The objective is to verify that the current density is constant across different device areas, and to estimate the power dissipated by Joule heating during programming. Here and throughout the manuscript, the bias is applied to the top electrode, and the bottom electrode is grounded (\textbf{Figure \ref{fig:Conduction} a}). The ferroelectric polarization in the thin films is verified using the PUND method \cite{bondurant1990ferroelectronic}. As shown in \textbf{Supplementary Figure S1}, complete switching is achieved at 1 V for a 2.5 kHz waveform. Devices are then programmed and measured in both polarization directions. The time elapsed between the programming and the read does not significantly affect the measurement, as shown in retention measurements in \textbf{Supplementary Figure S2}. The non-destructive read operation at low bias ($V\leq100$ mV) reveals a quasi-linear increase of the direct current as a function of the bias for both polarities, and in the High and Low Resistive States (HRS, LRS) (\textbf{Figure \ref{fig:Conduction} b}). As mentioned in the introduction, transport in the read regime is thermally activated. The experimental current-voltage characteristics at various temperatures are well described by the Ohmic conduction model, where the change in resistance is attributed to a change in the relative height of the conduction band (see \textbf{Supplementary Figure S3}). 

At higher bias, the conduction mechanisms differ in the positive and negative polarities (\textbf{Figure \ref{fig:Conduction} c}). When a positive bias $V>V_{tr}=140$ mV is applied on the top electrode, the current density \textit{J} varies with $V^{2}$, indicating a traps-filled limit (TFL) current \cite{chiu_review_2014}. In the linear regime, the density of injected carriers is smaller than the density of thermally generated free carriers. The injected carriers redistribute spatially in the oxide during a dielectric relaxation time $\tau_{r}$. Above the threshold $V_{tr}$, the space charge carriers filled all the traps in the dielectric oxide, leading to an increase in the free carrier density as the bias \textit{V} increases. In this TFL regime, the current density is 
\begin{equation}\label{TFL_equation}
J_{TFL}=\frac{8}{9}\mu\varepsilon\theta\frac{V^{2}}{d^{3}},
\end{equation} where $\mu$ is the electron mobility, $\varepsilon$ is the static dielectric constant, $\theta$ is the ratio of the free carrier density to total carrier density, and $d$ is the thickness of the insulator.
Instead, when a positive bias is applied to the bottom electrode, conduction is no longer limited by the bulk, but rather by the injection of carriers at the electrode. The experimental data are consistent with a Schottky emission mechanism: 

\begin{equation}\label{Schottky_equation}
J=\frac{120m^{*}}{m_{0}}{T}^{2}exp\left[ \frac{-q\left( \phi_{B}-\sqrt{qE/4\pi\varepsilon_{r}\varepsilon_{0}} \right)}{kT} \right],
\end{equation}

where $m_{0}$ is the free electron mass, $m^{*}$ refers to the effective electron mass in the oxide, $T$ is the temperature, $q\phi_{B}$ is the barrier height, $\varepsilon_{0}$ is the vacuum permittivity in vacuum, and $\varepsilon_{r}$ represents the dielectric constant.

 For both polarities, current-voltage characteristics were measured at various temperatures. As shown in \textbf{Supplementary Figure S4} and \textbf{Figure S5}, the analysis of these characteristics using the TFL and Schottky models consistently yields an activation energy of 0.28 eV (TFL regime) and a barrier height of 0.21 eV (Schottky emission regime). Although this apparent barrier is lower than the ideal Schottky barrier expected from band-alignment arguments, such reduced transport-extracted values are well documented in non-ideal Schottky contacts, where barrier inhomogeneity and defect-mediated low-energy conduction paths can dominate the measured current \cite{tungPhysicsChemistrySchottky2014}. Effective barriers of a few tenths of eV have been reported in HfO\textsubscript{2}-based stacks \cite{walczykPulseinducedLowpowerResistive2009, panHightemperatureConductionBehaviors2007}. The observed asymmetry originates from the presence of asymmetric electrodes (TiN on the top electrode and WO\textsubscript{x} on the bottom electrode). The conduction models presented in Figure \ref{fig:Conduction} were obtained by fitting the experimental \textit{I-V} for a 4 \textmu m\textsuperscript{2} weight. \textbf{Figure \ref{fig:Conduction} c)} further displays the \textit{J-V} characteristics for four devices of various dimensions, calculated by dividing \textit{I} by the nominal area. The DC \textit{J-V} characteristics overlap, confirming that the conduction scales well with the area.
 Figure \ref{fig:Conduction} c) reveals a spread in the current-density characteristics. As expected, the device-to-device variability increases as the device dimension decreases (see \textbf{Supplementary Figure S6} for a comparison of two area ranges).

From these equations, we derive the energy \textit{Q} dissipated by Joule heating when a programming pulse of amplitude \textit{V\textsubscript{write}} and duration \textit{t\textsubscript{write}} is applied: \textit{Q=IV\textsubscript{write}t\textsubscript{write}}. 
We estimate an upper bound for Joule heating, for 20 ns pulses and \textit{V\textsubscript{write}} comprised between $\pm$3 V. For a device of area 24 \textmu m\textsuperscript{2}, the current density \textit{J\textsubscript{max}} is the highest for \textit{V\textsubscript{write}}=-3 V, with \textit{J\textsubscript{max}}=$-2.2\cdot10^{6}$ A/m\textsuperscript{2}. The resulting upper bound for the programming energy is \textit{Q }= 3.1 pJ.

\subsection{Ferroelectric switching in scaled devices}

Instead, in this work, the laterally scaled resistive weights maintain their On/Off ratio as the pulse duration decreases. This is understood in terms of characteristic charging time of the oxide thin film, as discussed for example by Menzel \textit{et al.}\ \cite{menzel_ultimate_2019}. 
As an electric pulse is applied to a dielectric thin film sandwiched between two metallic electrodes, the equivalent circuit is a resistance \textit{R\textsubscript{dev}} in parallel with a capacitance \textit{C\textsubscript{dev}}, and in series with the electrode resistance \textit{R\textsubscript{el}}.

In ref.\ \cite{menzel_ultimate_2019}, the authors show for a similar geometry that the characteristic time constant $\tau$ is equal to:
\begin{equation}
    \tau=\frac{R_{dev}R_{el}}{R_{dev}+R_{el}}C_{dev} .
\end{equation}

Although \textit{R\textsubscript{dev}} and \textit{C\textsubscript{dev}} scale with the device area, it is not the case for the electrode resistance \textit{R\textsubscript{el}}, which is approximately 5 kOhms in our system. This value was measured after the electrical breakdown of a device. The self-charging time is thus expressed as a function of the dielectric constant \textepsilon\ and the film thickness \textit{d}:

\begin{equation}
    \tau=\frac{VR_{el}\varepsilon_{dev}A}{d(V+R_{el}J_{dev}A)}.
\end{equation}

Observing that the leakage current density \textit{J} is independent of device area, and that $R_{el}$ is constant, this expression predicts that \texttau\ decreases when the device area \textit{A} decreases. 

In order to estimate \texttau\, the capacitance of the devices is measured at 5 MHz for different devices. We verify that the capacitance at zero bias scales linearly with the area (\textbf{Figure \ref{fig:Conduction} d}), and that under the selected operational voltages C is independent of V. \textit{R\textsubscript{dev}} and \textit{J\textsubscript{dev}} vary with \textit{V}: using the results of \textbf{Figure \ref{fig:Conduction} c}, we estimate the dependence of \texttau  \ on the device area for different values of \textit{V}. \textbf{Figure \ref{fig:Conduction} e} shows that for devices larger than 100 \textmu m\textsuperscript{2}, \texttau\ compares with 20 ns. 
When $\tau$ exceeds the pulse duration, the effective potential across the ferroelectric thin film becomes smaller than the applied voltage, preventing programming and leading to a collapse of the On/Off ratio.

\begin{figure}[hbt!]
  \includegraphics[width=\linewidth]{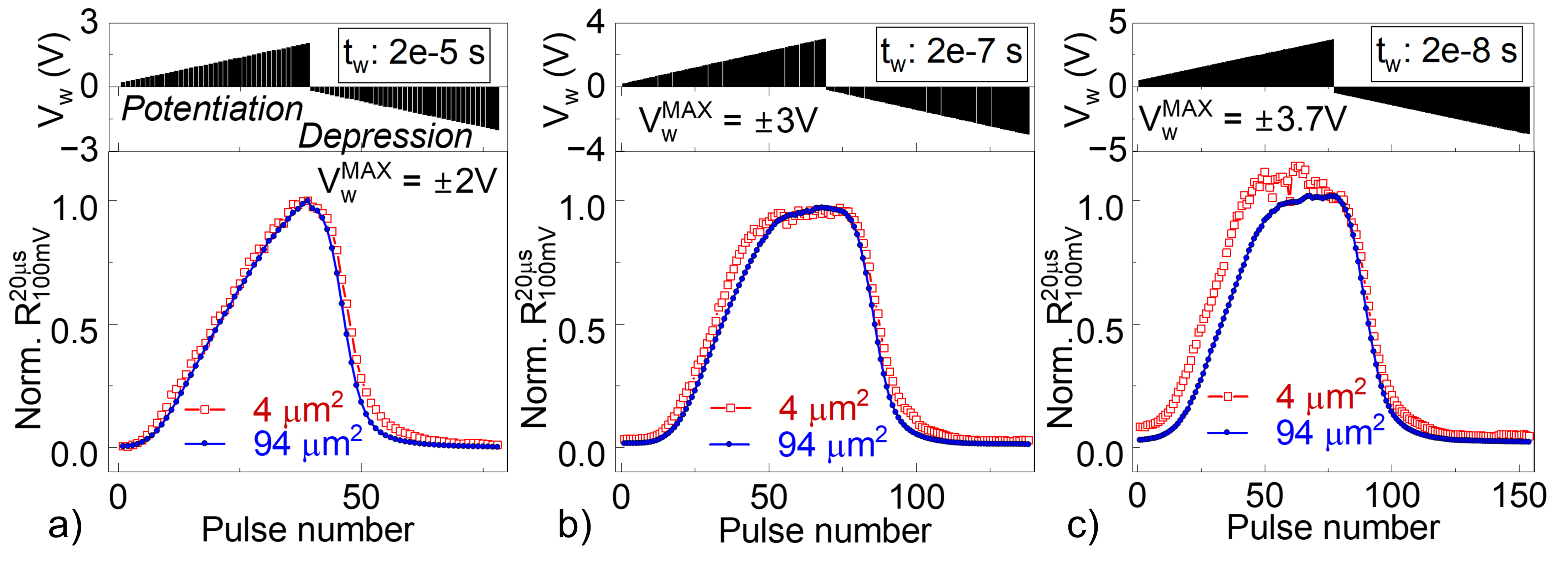}
  \caption{Potentiation and depression characteristics for two devices of area 4 \textmu m\textsuperscript{2} (red symbols) and 94 \textmu m\textsuperscript{2} (blue symbols). The weight is calculated by normalizing the read resistance (at\textit{V\textsubscript{read}} = 100 mV) by the dynamic range extracted from LTP/LTD experiments at \textit{t\textsubscript{write}} = 20 \textmu s. In a), \textit{t\textsubscript{write}} = 20 \textmu s and \textit{V\textsubscript{write}} varies between 2 V and -2 V. In b), \textit{t\textsubscript{write}} = 200 ns and \textit{V\textsubscript{write}} varies between 3 V and -3 V. In c), \textit{t\textsubscript{write}} = 20 ns and \textit{V\textsubscript{write}} varies between 3.75 V and -3.75 V. } 
  \label{fig:LTPLTD}
\end{figure}

To verify this statement, different devices with an area comprised between 94 and 1.4 \textmu m\textsuperscript{2} are measured under different pulse widths \textit{t\textsubscript{write}} of 20 ns, 200 ns and 20 \textmu s. After each programming pulse of amplitude \textit{V\textsubscript{write}} and duration \textit{t\textsubscript{write}}, the resistance is measured non-destructively at \textit{V\textsubscript{read}} = 100 mV. After each write-read sequence, \textit{V\textsubscript{write}} is gradually increased, as shown in the upper boxes of \textbf{Figure \ref{fig:LTPLTD}}.
Ferroelectric switching in polycrystalline materials generally occurs through the nucleation and anisotropic growth of individual domains. The creep velocity of domain wall motion is inversely proportional to the switching time and follows the empirical Merz law \cite{merz_domain_1954}:
\begin{equation}\label{Merz}
{t_{0}=t_{\infty} \exp\left( \frac{E_{act}}{E}\right)}.
\end{equation}
Consequently, programming the weights with shorter voltage pulses requires one to increase the maximal amplitude accordingly: from 2 V for 20 \textmu s, it is increased to 3 V for 200 ns pulses and 3.75 V for 20 ns pulses. The respective step sizes are 48 mV, 41 mV, and 42 mV. 
For each measurement, the read resistance \textit{R} is normalized by the dynamic range measured for the same device, operated under \textit{t\textsubscript{write}} = 20 \textmu s. 
A representative cycle is represented for two sizes and three pulse widths in \textbf{Figure \ref{fig:LTPLTD} a, b} and \textbf{c}. The behavior is reproducible, however, an abrupt change in resistance is occasionally observed on the smaller devices, as shown in \textbf{Supplementary Figure S7}. A possible explanation is that for micrometer-scale devices, the number of ferroelectric domains becomes discrete, leading to a stochastic behavior \cite{mulaosmanovic_switching_2017}. By defining the coercive voltage V\textsubscript{C} as the optima of the dR/dV derivatives of the LTP and LTD curves, we can verify that the thin films follow the Merz law (\textbf{Supplementary Figure S8}) \cite{merz_domain_1954}. The extrapolated activation voltages are Va=-2.9 V for negative bias and Va=5.3 V for positive bias. The activation fields, estimated for a thickness of 5 nm, are 5.8 and 11 MV.cm\textsuperscript{-1}, respectively. These values are consistent with previous reports demonstrating that the switching dynamics in ultra-thin HfZrO\textsubscript{4} were described by the Nucleation Limited Switching framework \cite{kondratyuk2022polarization}.
Comparing \textbf{Figure \ref{fig:LTPLTD} a, b} and \textbf{c}, we observe that both for a 4 \textmu m\textsuperscript{2} and a 94 \textmu m\textsuperscript{2} device, there is no degradation of the On/Off ratio upon decreasing \textit{t\textsubscript{write}}. In contrast, in previous work on large area (10000 \textmu m\textsuperscript{2}) resistive weights revealed the collapse of the On/Off ratio as the programming pulse width decreases: the On/Off ratio was reduced by 76\% for 20 ns pulses compared to 20 \textmu s pulses \cite{begon-lours_back-end--line_2024}. These results confirm that the self-loading time of the devices in this work is shorter than 20 ns.

\FloatBarrier

\subsection{Asymmetric resistive switching in hafnium zirconate / tungsten oxide bilayers.}

\begin{figure}[hbt!]
    \centering
  \includegraphics[width=1\linewidth]{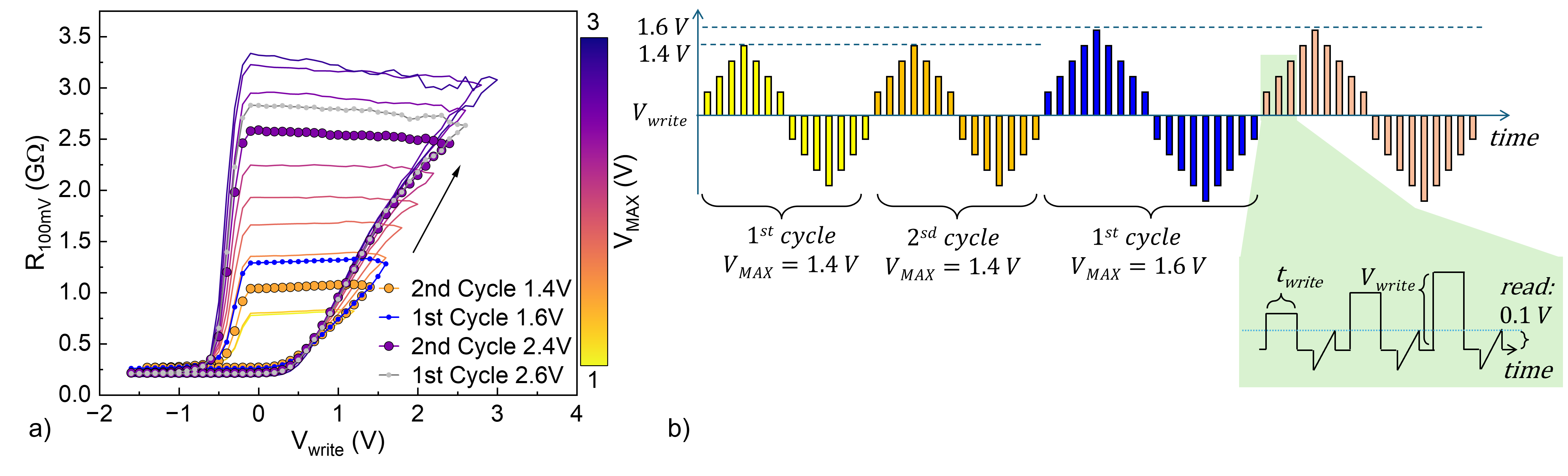}
  \caption{a) Steady-state Resistance (R\textsubscript{100mV}) after DC pulses (40 ms) of amplitude (V\textsubscript{write}) on a 24 \textmu m\textsuperscript{2} device. Cycles are performed twice before increasing the maximum amplitude (V\textsubscript{MAX}), and the second cycles are presented as lines. The color bar indicates the maximum amplitude (V\textsubscript{MAX}) of each cycle. Comparing the second cycle with the first cycle at increased amplitude (scatter data), we observe that it follows the path of the previous loop before accumulating further damage caused by the high field applied for long time periods by the DC pseudo-pulses. This leads to an increase in the coercive field and electrostatic charging. b) Sketch of the measurement scheme representing two cycles at the same V\textsubscript{MAX} (for example 1.4 V) followed by two cycles at a higher \textsubscript{MAX} (here 1.6 V). A zoom in the measurement scheme illustrates that between each programming pulse, the resistance is measured at V\textsubscript{read} = 100 mV.}
  \label{fig:RVLoops} 
\end{figure}

To get insights into the mechanisms leading to the asymmetry in the LTP and LTD branches, complete hysteresis loops are measured. Write-read sequences are applied. For successive programming pulses, the voltage \textit{V\textsubscript{write}} is gradually increased until \textit{V\textsubscript{MAX}}, and then gradually decreased from \textit{V\textsubscript{write}} = \textit{V\textsubscript{MAX}} to \textit{V\textsubscript{write}} = -1.6 V, at which point the polarization saturates in the LRS. The bias is applied in DC mode, corresponding to a long integration time of 40 ms. After each write pulse, the resistance is measured at \textit{V\textsubscript{read}} = 100 mV. For each \textit{V\textsubscript{MAX}} increase, two hysteresis loops are measured. For the first two loops, \textit{V\textsubscript{MAX}} is set to 1 V (yellow line in Figure \ref{fig:RVLoops}). Then, the experiment is repeated with a higher \textit{V\textsubscript{MAX}} up to 3 V (dark blue line in Figure \ref{fig:RVLoops}).

Figure \ref{fig:RVLoops} reveals four observations. First, that increasing \textit{V\textsubscript{MAX}} from a pristine condition results in a steeper weight update. The resistance measured at a given \textit{V\textsubscript{write}} increases for a loop taken after a higher \textit{V\textsubscript{MAX}} is applied.
Second, we find that, in contrast, the potentiation and depression data for two consecutive loops after the same maximal voltage \textit{V\textsubscript{MAX}} overlap. This is illustrated by the scattered data (filled circles), for example, after \textit{V\textsubscript{MAX}} = 1.4 V (blue and orange symbols in \textbf{Figure \ref{fig:RVLoops}}), and after \textit{V\textsubscript{MAX}} = 2.4 V (grey and purple symbols in \textbf{Figure \ref{fig:RVLoops}}). Consistently, the read resistance measured upon the application of pulses of decreasing amplitude (horizontal branches of the hysteresis loops) is constant. It indicates that increasing the V\textsubscript{MAX} is responsible for the irreversible changes in the device properties. 
Third, we observe that the change in resistance after increasing \textit{V\textsubscript{MAX}} is more pronounced in the High Resistive State (HRS) than in the Low Resistive State (LRS).
Finally, the LRS saturates for \textit{V\textsubscript{write}} below -0.8 V: this is expected as it corresponds to a situation where all the ferroelectric domains have switched and are aligned along the same direction. However, no saturation is observed for the HRS, even for \textit{V\textsubscript{write}} as high as 3 V. For such high voltage, the slight increase of the read resistance upon decreasing \textit{V\textsubscript{write}} indicates that, under long programming pulses and large bias, other mechanisms than ferroelectric switching are involved. Such mechanisms could be reversible ion motion \cite{nukala_reversible_2021}, redox reactions between hafnia and WO\textsubscript{x} \cite{halter_multi-timescale_2023}, or charge trapping \cite{barbot_interplay_2022}. 
In the literature, it is generally reported that such operating conditions are detrimental to the endurance \cite{lenox2024impact}. Consistently, fatigue measurements show a stable behavior for a 1.5 V waveform, whereas it exhibits a change in the polarization after 10\textsuperscript{4} cycles when the amplitude is increased to 2 V (\textbf{Supplementary Figure S9}). The endurance of the devices is good, without dielectric breakdown after 10\textsuperscript{10} cycles.
Such an observation further motivates the use of short pulses for programming the synaptic weights.

\subsection{Weight update rule in ferroelectric resistive weights}

\begin{figure}
    \centering
  \includegraphics[width=0.5\linewidth]{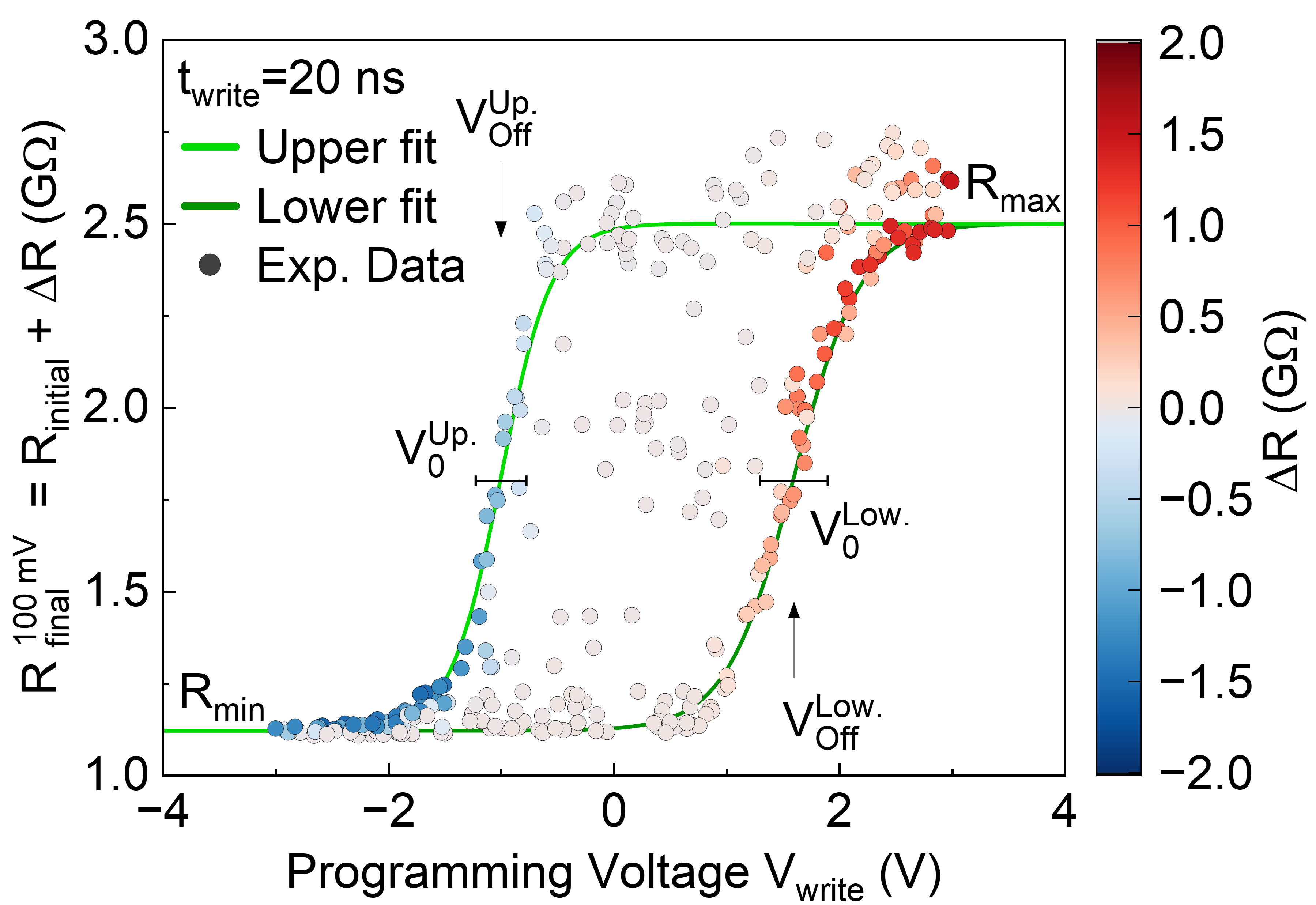}
  \caption{After a programming pulse of amplitude V\textsubscript{write} and duration t\textsubscript{write} = 20 ns, the resistance R\textsubscript{final} of the weight is measured at V\textsubscript{read} = 100 mV. The color bar (\textDelta R) represents the difference in resistance before and after the programming pulse \textDelta R=R\textsubscript{final}-R\textsubscript{initial}. The data envelope is fitted with two tanh functions.}
  \label{fig:RandomPulses}
\end{figure}

To simulate a training or weight transfer operation, we perform 300 independent measurements using a write-read sequence on a device. The pulses have a random amplitude (between -3 and +3V) at a fixed width (20 ns). The devices are not reset between the random writes. By considering the pulse order, we record the initial (\textit{R\textsubscript{initial}}) and final (\textit{R\textsubscript{final}}) resistance states achieved by a given write voltage (\textit{V\textsubscript{write}}), and how the resistance has changed (\textit{\textDelta R = R\textsubscript{final}-R\textsubscript{initial}}) (Figure \ref{fig:RandomPulses}). This sequencing leads to a random combination of initial device states and amplitude for the write pulse. We note that in contrast with Figure \ref{fig:RVLoops}, the On/Off ratio (ratio between the maximal and minimal resistance) is moderate, consistent with a regime dominated by ferroelectric switching.    

We observe that the data points are all located inside a S shape envelope, which is characteristic of ferroelectric polarization switching. Inspired by the Preisach model for ferroelectric polarization \cite{bartic2001preisach}, we use a hyperbolic tangent function \textit{f\textsubscript{upper}}, resp.\ \textit{f\textsubscript{lower}} to fit the upper (resp.\ lower) branch of the envelope:

\begin{equation}\label{eq:6}
f: V \mapsto   R_{\mathrm{off}} + R_{s}\tanh\left(\frac{V - V_{\mathrm{off}}}{V_{0}}\right),
\end{equation}

where:

\begin{equation}\label{eq:7}
R_{\mathrm{off}} = \frac{R_{\max} + R_{\min}}{2} {,} \quad
R_{s} = \frac{R_{\max} - R_{\min}}{2}.
\end{equation}
Here, $R_{\mathrm{\min}}$ and $R_{\mathrm{\max}}$ denote the physical saturation limits of the device, $R_{\mathrm{off}}$ represents the midpoint between these two resistance states, and $R_{s}$ determines the amplitude of the switching window. The parameter $V_{\mathrm{off}}$ sets the voltage at which the transition is centered, while $V_{0}$ controls its steepness. The resulting parameter values are reported in Table \ref{tab:tanh_params}. 

\begin{table}[t]
\centering
\caption{Fitted parameters for the lower and upper switching envelopes.}
\begin{tabular}{c|c|c|c|c}
\hline
Envelope & $R_{\min}$ [$\Omega$] & $R_{\max}$ [$\Omega$] & $V_{0}$ [V] & $V_{\text{off}}$ [V] \\
\hline
Lower & $1.1\times10^{9}$ & $2.5\times10^{9}$ & $0.60$ & $1.6$ \\
Upper & $1.1\times10^{9}$ & $2.5\times10^{9}$ & $0.45$ & $-1.0$ \\
\hline
\end{tabular}
\label{tab:tanh_params}
\end{table}

Notably, any change in resistance of the device appears to be solely located in the envelope of the data set: all datapoints within the hysteresis loop do not show any change in conductance (gray color). The variance in resistance after non-switching pulses is found to be below 100 M\textOmega\ (\textbf{Supplementary Figure S11}). This indicates that: 1) the conductance is agnostic on the original state as long as we apply enough field to go to a higher or a lower resistive state, and 2) the device conductance only changes through a narrow field regime that can be exploited to program precise conductance levels.
We propose the introduction of a weight update rule: 
to program a specific weight, the system does not need to measure or know the exact actual weight value, just the direction required for the switch (towards a higher or lower resistance value). This can be achieved by either storing the weight during one time step and using it to choose the appropriate envelope, or applying a full reset to either polarity and using the corresponding envelope. The new specific weight can then be programmed by applying a single 20-nanosecond electrical pulse. We formalize the following weight update rule based on the functions derived from equation \ref{eq:6} and table \ref{tab:tanh_params}:

\begin{equation}
\label{eq:updaterule}
R_{final} = \begin{cases*}
      f_{upper}(V_{write}), & if $V_{write} < 0 \ and \ R_{initial} > f_{upper}(V_{write})$,\\
      f_{lower}(V_{write}), & if $V_{write} > 0 \ and \ R_{initial} < f_{lower}(V_{write})$, \\
      R_{initial},          & otherwise.
\end{cases*}
\end{equation}

The device switching behavior can be accurately described using a model based on $\tanh$ using a small set of physically significant parameters $(R_{\min}, R_{\max}, V_{0}, V_{\text{off}})$. A statistical analysis on the change in resistance after a write pulse in points found in the envelope reports a deviation between the experimental and expected \textDelta R values below 300 M\textOmega\ (\textbf{Supplementary Figure S12}).

\subsection{Application: Voltage-Driven Synaptic Plasticity}

\begin{figure} [h]
  \includegraphics[width=\linewidth]{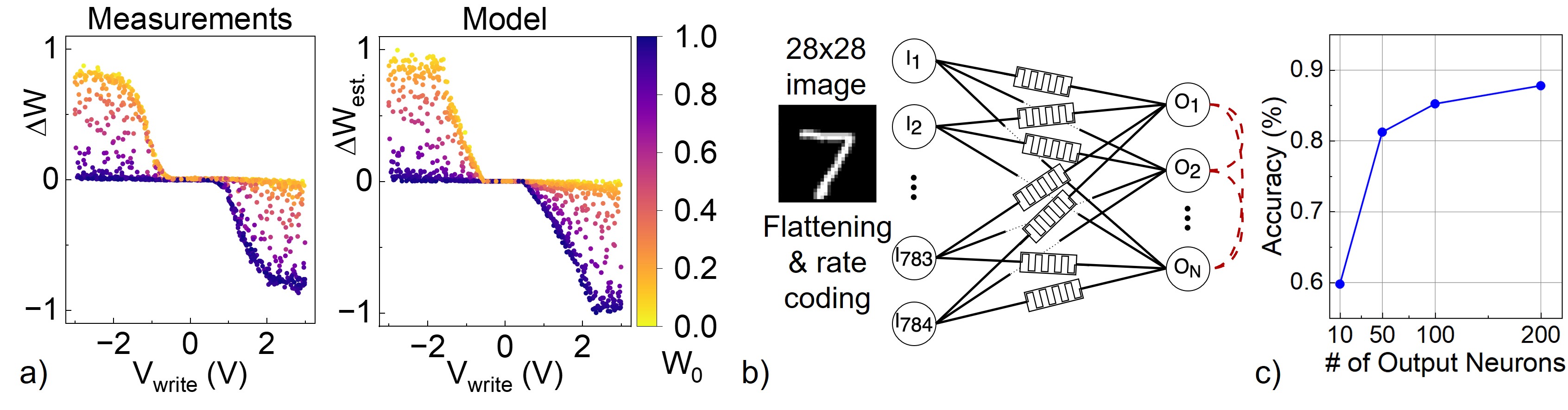}
  \caption{Voltage-dependent synaptic plasticity-based online learning. a) Normalized weight-update magnitude is plotted against programming voltage and initial weight, with both the measured data and model response displayed. b) The schematic shows the spiking neural network, in which a 28×28 grayscale image from the MNIST database is flattened and provided to 784 input neurons, which are densely connected to N output neurons through memristive synapses updated by the VDSP learning rule. c) Network's classification performance plotted as a function of the number of output neurons. }
  \label{fig:VDSP}
\end{figure}

As an outlook, an application is proposed for the studied synaptic weights. In refs.\ \cite{garg2022voltage, gargUnsupervisedLocalLearning2026}, N.\ Garg et al.\ proposed Voltage-dependent synaptic plasticity (VDSP) for circuit implementation of online learning with memristive synapses. We verify that operating the scaled ferroelectric weights with short pulses allows energy gains without compromising application-level performance. This Hebbian plasticity rule is based on the following principle: at the time of a post-synaptic neuron spike on a given synapse, the membrane potential of the pre-synaptic neuron is measured. If the membrane potential is low, the pre-synaptic neuron just fired, \textit{i.e.}, the temporal correlation is positive and therefore the synapse should be potentiated. In contrast, if the membrane potential is high, it means that the pre-synaptic neuron is about to fire: there is an anti-correlation. In this case, the synapse should be depressed. This bio-inspired plasticity rule is implemented by directly converting the neuron membrane potential into a programming pulse. Therefore, this simplifies the implementation of spike timing-dependent synaptic plasticity (STDP) by avoiding complex pulse shaping and overlapping circuits, and fully leverages the analog programming of memristors. To predict the learning ability of the circuit, the simulation framework described in ref.\ \cite{gargUnsupervisedLocalLearning2026} is used. It is not based on the description of the final weight as a function of the voltage amplitude as in eq.\ \ref{eq:updaterule}. Instead, it is based on a device model describing the change in weight as a function of the programming voltage and of the initial weight. From the experimental data in Figure \ref{fig:RandomPulses}, we derive the required representation: the weight update as a function of the write amplitude (see Figure \ref{fig:VDSP} a). The equation used to fit the experimental data, taken from \cite{gargUnsupervisedLocalLearning2026}, is described in the methods section. The model (Figure \ref{fig:VDSP} b)) fits the experimental data using the parameters $\gamma_{p} = 1.62$, $\gamma_{d} = 1.79$, $\alpha_{p} = 0.67$, $\alpha_{d} = 0.38$, $\theta_{p} = 0.55$, and $\theta_{d} = 0.47$. Here, $\alpha{p}$ and $\alpha{d}$ denote the coefficients of fitting of exponential curvature associated with potentiation
and depression with respect to the bias, respectively; and $\theta{p}$ and $\theta{d}$ are the corresponding threshold
voltages that govern memristive device switching during potentiation and depression. $\gamma{p}$ and $\gamma{d}$ quantify the potentiation and depression non-linearity with respect to the weight value (see Experimental Section 4.3 ). The same method as in the original paper is used to simulate the accuracy as a function of the number of output neurons for the MNIST classification task. The results show that a similar accuracy (87.88\% for 200 output neurons) is obtained when reducing the pulse duration by three orders of magnitude, and consequently reducing the energy cost during the learning phase.

The reported classification accuracy is consistent with prior work on unsupervised spiking neural networks trained using STDP. For instance, fully software-based unsupervised STDP applied to MNIST achieves an accuracy of 82.9\% for a single-layer, fully connected network with 100 output neurons \cite{diehl2015unsupervised}. Comparable performance levels have also been reported for STDP-trained memristive spiking neural networks \cite{querlioz2013immunity, boybat2018neuromorphic}. The gap with respect to supervised fully connected networks primarily reflects the limitations of single-layer architectures trained with unsupervised learning, rather than constraints imposed by the device model or the proposed update rule. Notably, in prior work, the same voltage-dependent plasticity rule, when applied to spiking convolutional neural networks, achieved accuracies exceeding 98\% on MNIST \cite{goupy2023unsupervised}, benefiting from the improved spatial feature extraction enabled by convolutional topologies. Building on these results, future work will investigate the deployment of the proposed learning rule in deeper and convolutional spiking architectures, as well as the integration of a third modulatory factor to enable task-driven or reward-modulated learning.

\FloatBarrier

\section{Conclusion}

In this work, we addressed the challenge of reducing the energy cost associated with the training of neuromorphic systems based on memristive technologies. We fabricated ferroelectric resistive devices based on hafnia/zirconia nanolaminates using a fully Back-End-Of-Line compatible process. By laterally scaling the device area below 100 \textmu m\textsuperscript{2}, we demonstrated that the self-loading time becomes sufficiently short to enable reliable 20-nanosecond programming pulses, thus achieving ultrafast operation without compromising the On/Off ratio or device endurance. The programming pulses in this regime only dissipate a maximum of 3 pJ.
We experimentally determined the weight update rule for these devices under 20-nanosecond pulses across different voltage amplitudes and initial conductance states. The results reveal that the final conductance is governed exclusively by the pulse amplitude and remains independent of the initial weight. This behavior defines a deterministic and amplitude-controlled learning mechanism, simplifying the implementation of training and weight transfer operations in neuromorphic circuits.
Overall, this study establishes a clear design pathway toward high-speed, low-energy neuromorphic hardware by linking ferroelectric material scaling to circuit-level performance. The demonstrated 20-nanosecond programming regime, combined with a predictable and amplitude-dependent update rule, paves the way for efficient on-chip learning in CMOS-compatible neuromorphic accelerators.


\section{Experimental Section}
\subsection{Fabrication}

The substrate is a 2 \textmu m thick thermal SiO\textsubscript{2} on Si chip. A W thin film (50 nm) was sputtered prior to the functional stack, identical to that of ref.\ \cite{begon-lours_back-end--line_2024}. Using an Oxford Instruments plasma enhanced atomic layer deposition (PEALD) system, 20 nm of TiN was deposited at 300\textdegree C, then 45 cycles of WO\textsubscript{x} at 360\textdegree C. The growth rate was determined by ellipsometry and is 0.495 A/cycle. Afterwards, a 5 nm [HfO\textsubscript{2}/ZrO\textsubscript{2}] nanolaminate is deposited at 300\textdegree C, at 0.5 nm per layer. It consists of five supercycles of HfO\textsubscript{2} and ZrO\textsubscript{2}, each supercycle comprising five cycles with tetrakis (ethylmethylamino) hafnium (IV) (TEMAH) and O\textsubscript{2} plasma, growth rate 1.05 A/cycle, and ten cycles with bis (methylcyclopentadienyl) (methyl) (methoxy) zirconium (IV) (ZRCMMM) and O\textsubscript{2} plasma, growth rate 0.548 A/cycle. Previous analysis on films deposited in the same conditions but as a solid solution indicated a composition of Hf\textsubscript{0.57}Zr\textsubscript{0.43}O\textsubscript{2} \cite{oconnor_stabilization_2018}. The X-Ray Reflectivity spectra of a HZO solid solution (25 supercycles comprising 1 cycle of TEMAH and 2 cycles ZRCMMM) and of a (HfO\textsubscript{2}/ZrO\textsubscript{2}) stack (5 supercycles comprising 5 cycles of TEMAH and 10 cycles ZRCMMM) exhibit comparable fringes (\textbf{Supplementary Figure S9}), indicating that the total thickness of these two thin films is similar. However, the electrical properties of thin films fabricated with one or five cycles per supercycle are generally different \cite{lehninger2025advancing, begon-lours_back-end--line_2024}.
Ten nanometers of TiN were deposited in situ. Crystallization was performed with a millisecond flash lamp anneal: the sample was preheated at 375\textdegree C, followed by a 20 ms pulse of 90 J/cm\textsuperscript{2}.
A 50 nm thick W metal electrode was then sputtered. The top electrode was defined by e-beam lithography and reactive ion etching (RIE). The bottom electrode was defined by optical lithography and ion-beam etching of the HfO\textsubscript{2}/ZrO\textsubscript{2}, WO\textsubscript{x}, TiN, and W layers. 
A 100 nm thick SiO\textsubscript{2} passivation layer was then sputtered. Vias to the device’s top and bottom electrode contacts were defined by e-beam lithography. The SiO\textsubscript{2} layer was etched by RIE, and then the HfO\textsubscript{2}/ZrO\textsubscript{2} and the WO\textsubscript{x} were etched by ion beam etching. 100 nm of W was then sputtered. The metal lines were then defined by optical lithography and RIE.

\subsection{Electrical characterization}

All pulsed electrical measurements were carried out using an Agilent B1500A semiconductor parameter analyzer. No wake-up procedure is needed to observe resistive switching. For each new, pristine device investigated, the device was first measured with 5 DC cycles at low voltage (200 mV). Write pulses V\textsubscript{write} were applied to the terminal connected to the top electrode of the devices, while the bottom electrode was shorted to ground through the second terminal.
Fatigue measurements confirm that unless the device shows stable behavior up to 10\textsuperscript{4} cycles under high bias, and up to 10\textsuperscript{8} cycles under moderate bias (\textbf{Supplementary Figure S10}).
The pulses were generated using the waveform generator/fast measurement unit (WGFMU) in combination with the remote-sense and switch unit (RSU) module, which is integrated into the B1500A system. The differential read resistance (R\textsubscript{100mV}) was extracted by performing a voltage sweep from –100 mV to +100 mV between top and bottom electrodes of the Metal-Ferroelectric-Metal resistive devices. The resistance was calculated as the average value at the two extreme bias points (-100 and +100 mV). For R\textsubscript{100mV} versus V\textsubscript{write} characterization, alternating write and read steps were executed while progressively ramping V\textsubscript{write} up and down to trace a full hysteresis loop starting at 0 V.
In the LTP/LTD measurement, a reset conditioning step of 1000 cycles with an amplitude of $\pm$1.5 V at 1 kHz was applied in between experiments, before modifying the write pulse amplitude or width. The objective of the reset conditioning sequence is to provide the same starting point for each device, at a controlled time before the measurement. 
For the random-pulse VDSP experiment, the same write and read parameters were used, but instead of tracing the hysteresis loops with increasing voltage, random positive or negative pulses were applied. A sequence of 300 writes and reads was carried out. Capacitance measurements were carried out using the Multi-Frequency Capacitance Measurement Unit (MFCMU) module of the B1500A. The capacitance was measured in devices ranging from 94 to 1.4 \textmu m\textsuperscript{2} with an amplitude of $\pm$500 mV at the highest possible frequency (5 MHz), with a step size of 5 mV and a small-signal AC amplitude of 30 mV. Ferroelectric and fatigue measurements were carried out with an aixACCT Systems TF Analyser 3000 (see Supplementary Information S1 and S6).

\subsection{VDSP and SNN simulations}

To evaluate learning performance, spiking neural network (SNN) simulations were performed on the MNIST handwritten digit dataset. The methodology for the simulation is described in ref. \cite{gargUnsupervisedLocalLearning2026} and summarized here: input images were encoded as constant currents and converted into spike trains by an encoding layer of leaky integrate-and-fire neurons, with firing rates proportional to pixel intensities. These spikes are weighted by the memristive synapses and integrated by neurons in the output layer. The output neurons are connected through a winner-take-all topology, such that only a single neuron is active at a given time, corresponding to the network’s decision. The neuron model, training and evaluation procedure, and hyperparameters follow standard implementations used in prior work. Training was performed in an unsupervised manner without using labels with the 60,000 samples from the training subset of the MNIST dataset. After convergence, the synaptic weights were fixed, and the training dataset was reused to assign a class label to each output neuron. The trained network was then evaluated on the MNIST test set to compute the recognition accuracy.

The weight update, denoted by $\Delta W$, is modeled as the product of a switching rate function $f$, which depends on the applied voltage $v$, and a window function $g$, which depends on the weight $W$:
\begin{equation}
    \Delta W = f(v) \cdot g(W)
\end{equation}

\begin{equation}
    f(v) = \begin{cases} 
              e^{-\alpha_p \cdot (v - \theta_p)} - 1 & \text{if } v < \theta_p \\
              e^{\alpha_d \cdot (v - \theta_d)} - 1 & \text{if } v > \theta_d 
           \end{cases}
    \label{eq:f_of_v}
\end{equation}

Here, $\alpha_p$ and $\alpha_d$ denote the coefficients of fitting of exponential curvature associated with potentiation and depression, respectively; $v$ represents the applied voltage; and $\theta_p$ and $\theta_d$ are the corresponding threshold voltages that govern memristive device switching during potentiation and depression. 

\begin{equation}
    g(W) = \begin{cases}
        (1 - w)^{\gamma_p} & \text{if } v < \theta_p \\
        w^{\gamma_d} & \text{if } v > \theta_d
    \end{cases}
    \label{eq:g_of_W}
\end{equation}
The window function characterizes how the weight update depends on the initial state ($w$) and is responsible for the multiplicative behavior observed in successive switching events. $\gamma_p$ and $\gamma_d$ are the non-linear fitting parameters for potentiation and depression. The weights are clipped between 0 and 1 after update to avoid numerical error. 

A scaling factor ($sf$) is used to tune the actual voltage applied to the memristor so that it matches the operational range of the memristive device. This factor essentially translates the computing signals (in the form of membrane potential) into the appropriate voltage levels that a memristor requires for switching. The relationship between these parameters can be expressed as follows:

\begin{equation}
    V_{prog} = Vmem \cdot sf \cdot \theta
    \label{eq:sf}
\end{equation}

In this equation, the programming voltage ($V_{prog}$) is the product of the neuron membrane potential ($V_{mem}$), the memristor's fitted threshold value ($\theta$), and the scaling factor ($sf$). The $sf$ plays a similar role to the learning rate in ANNs as it regulates the degree of weight change. Only the scaling factors for LTP and LTD were optimized through grid search ($sf_p$, $sf_{d}$) for different network sizes. For networks with 10 output neurons, the scaling factors were set to $sf_p = 1.014$ and $sf_d = 1.30$. For 50 and 100 output neurons, $sf_p = 1.04$ and $sf_d = 1.30$ were used. For 200 output neurons, the values were $sf_p = 1.0725$ and $sf_d = 1.375$.

\medskip


\medskip
\textbf{Acknowledgements} \par 
We thank the Binning and Rohrer Nanotechnology Center, S.\ R.\ Mamidala, D.\ F.\ Falcone, U.\ Drechsler, A.\ Olziersky, and S.\ Reidt at IBM. Research funded by SNSF ROSUBIO \#218438, SERI SwissChips, EU Horizon.2.4 ViTFOX \#101194368, and EU Horizon.1.2 TOPOCOM \#101119608.

\textbf{Data Availability Statement} \par
The data that support the findings of this study are available from the corresponding author upon reasonable request.
%

\bibliographystyle{unsrt}
\bibliography{references,references-2}

\pagebreak








\end{document}